# Effect of layer thickness on structural, morphological and superconducting properties of Nb$_3$Sn films fabricated by multilayer sequential sputtering

M N Sayeed[1,2], U Pudasaini[3], C E Reece[4], G V Eremeev[4], H E Elsayed-Ali[1,2]

[1]Department of Electrical and Computer Engineering, Old Dominion University, Norfolk, VA 23529, USA

[2]Applied Research Center, 12050 Jefferson Avenue, Newport News, VA 23606, USA

[3]Applied Science Department, The College of William and Mary, Williamsburg, VA 23185, USA

[4]Thomas Jefferson National Accelerator Facility, Newport News, VA 23606, USA

E-mail: msaye004@odu.edu

**Abstract**.

Superconducting Nb$_3$Sn films can be synthesized by controlling the atomic concentration of Sn. Multilayer sequential sputtering of Nb and Sn thin films followed by high temperature annealing is considered as a method to fabricate Nb$_3$Sn films, where the Sn composition of the deposited films can be controlled by the thickness of alternating Nb and Sn layers. We report on the structural, morphological and superconducting properties of Nb$_3$Sn films fabricated by multilayer sequential sputtering of Nb and Sn films on sapphire substrates followed by annealing at 950 °C for 3 h. We have investigated the effect of Nb and Sn layer thickness and Nb:Sn ratio on the properties of the Nb$_3$Sn films. The crystal structure, surface morphology, surface topography, and film composition were characterized by X-ray diffraction (XRD), scanning electron microscopy (SEM), atomic force microscopy (AFM), and energy dispersive X-ray spectroscopy (EDS). The results showed Sn loss from the surface due to evaporation during annealing. Superconducting Nb$_3$Sn films of critical temperature up to 17.93 K were fabricated.

## 1. Introduction

Nb$_3$Sn is one of the type II superconductors that has wide applications from DC high-field magnets to radiofrequency cavities [1–5]. Magnetron sputtering is considered as one of the fabrication methods used to synthesize Nb$_3$Sn films. Nb$_3$Sn can be fabricated either from a stoichiometric Nb$_3$Sn single target [6, 7], or from sputtering of Nb and Sn followed by annealing [8]. Deposited films should maintain a controlled Sn composition range with atomic composition of 17–26% to obtain superconducting Nb$_3$Sn films [3]. For multilayer growth, the atomic Sn concentration of the films can be controlled by varying the thickness of the Nb and Sn layers.

We report on the structural, morphological and electrical properties of Nb$_3$Sn films fabricated by multilayer sequential sputtering of Nb and Sn films on sapphire substrates followed by annealing at 950 °C for 3 h. The temperature and time were selected based on previously reported work on Nb$_3$Sn fabrication [8]. The film properties were characterized for different layer thicknesses of Nb and Sn multilayers. The role of layer thickness was studied by varying the thickness of the Nb layers while keeping the Sn layer thickness constant, which adjusts the stoichiometry and by varying the thickness of both Nb and Sn layers while keeping the Nb:Sn thickness ratio constant.

## 2. Experimental details

### 2.1. Fabrication

Nb$_3$Sn films were grown by annealing sequentially-sputtered multilayers of Nb and Sn. Some details on the deposition chamber, substrate, and target materials were described elsewhere [8]. Multiple layers of Nb and Sn with different thicknesses were deposited sequentially with the first layer Sn and the final layer Nb. The final Nb layer minimizes Sn loss during post-deposition annealing. Films were deposited in a high vacuum chamber at $4 \times 10^{-3}$ mbar Ar sputtering gas (99.999% purity) with a flow rate of 20 SCCM. The substrate holder was rotated at 30 rpm throughout the sputtering process to maintain a homogenous deposition. Information about the film thickness is shown in Table 1. In the study of film stoichiometry, we chose a constant thickness of 10 nm of Sn and four thicknesses (10, 20, 30 and 40 nm) of Nb layers. These four conditions are referred to as conditions 1–4 in our results. The total thickness for all four conditions was kept constant (~900 nm) by varying the number of layers. In another set of films, we kept the Nb-to-Sn thickness ratio 2:1. For all films on these conditions, the total thickness was ~1.2 µm. These four conditions will be referred as conditions 5–8 in our results. All films were annealed in a separate vacuum furnace at 950 °C for 3 h with a temperature ramp rate of 12 °C/min.

**Table 1.** Deposition conditions of the films.

|  | Condition | Nb layer thickness (nm) | Sn layer thickness (nm) | Nb:Sn | Total thickness (nm) |
|---|---|---|---|---|---|
| Films with different stoichiometry | 1 | 10 | 10 | 1 | 900 |
|  | 2 | 20 | 10 | 2 | 900 |
|  | 3 | 30 | 10 | 3 | 900 |
|  | 4 | 40 | 10 | 4 | 900 |
| Films with fixed stoichiometry | 5 | 10 | 5 | 2 | 1200 |
|  | 6 | 20 | 10 | 2 | 1200 |
|  | 7 | 50 | 25 | 2 | 1200 |
|  | 8 | 200 | 100 | 2 | 1200 |

*2.2. Characterization*

X-ray diffraction (XRD) patterns of the films were obtained from Rigaku Miniflex II X-ray diffractometer with Cu-Kα radiation. The crystallite sizes of the films were measured from the peak broadening using Scherrer's equation [9]. The film microstructure was observed by field-emission scanning electron microscopic (SEM) images (FESEM S-4700, Hitachi, Japan). A Noran 6 energy dispersive X-ray spectroscopy (EDS) detector connected to a Jeol JSM 6060 LV scanning electron microscope system was used to quantitatively measure the surface composition. EDS was performed using 15 kV accelerating voltage on a surface covering 1.2 mm$^2$ area from each condition and the composition was measured on 5 different areas. The surface roughness of the samples were measured from the AFM images obtained by a Digital Instrument Dimension 3100 Atomic Force Microscope (AFM) in tapping mode. Root mean square (RMS) roughness was obtained over 5 × 5 µm scan area. Superconducting $T_c$'s were measured by four-point probe resistance measurement down to cryogenic temperature. $T_c$ was calculated from the mean value of $T^{90}$ and $T^{10}$ (temperature when the resistance is 90% and 10% of the final resistance before transition respectively) and the superconducting transition width $\Delta T_c$ was calculated from the difference of $T^{90}$ and $T^{10}$.

## 3. Results

### 3.1. Film with different stoichiometry

#### 3.1.1. Structural properties.

The Sn composition of as-deposited and annealed films are shown in Table 2. The Sn composition was reduced with increasing Nb layer thickness on as-deposited films. About 43.6% Sn was observed on films with condition 1, where both Nb and Sn layer thickness was maintained at 10 nm. All annealed films showed Sn loss after annealing. Large amount of Sn loss was observed on films with condition 1. The Sn composition changed from ~43.6% to ~23.8%. This large amount of Sn loss occurred due to sublimation of Sn during annealing. All annealed films showed Sn composition ~20–23%. The composition of the as-deposited films was ~16% in condition 4, however, the annealed film had ~20% Sn. Annealed film showed more Sn due to the uniform distribution of Sn throughout the film after annealing.

**Table 2.** Summary of structural, morphological, and superconducting data.

| Condition | At. % Sn as-deposited | At. % Sn annealed | RMS Roughness (nm) | $T_c$ (K) | $\Delta T_c$ (K) |
|---|---|---|---|---|---|
| 1 | 43.6 | 23.8 | 66.2±24.3 | 17.93 | 0.02 |
| 2 | 28.4 | 23.0 | 31.8±3.2 | 17.84 | 0.03 |
| 3 | 20.3 | 21.4 | 31.6±0.9 | 17.56 | 0.09 |
| 4 | 16.0 | 20.2 | 26.8±1.2 | 17.54 | 0.11 |
| 5 | 29.5 | 23.1 | 28.2±2.9 | 17.82 | 0.02 |
| 6 | 28.9 | 21.7 | 31.5±1.2 | 17.83 | 0.01 |
| 7 | 27.2 | 22.4 | 41.0±5.5 | 17.83 | 0.02 |
| 8 | 20.0 | 21.4 | 64.7±3.2 | 17.84 | 0.03 |

Figure 1 shows the SEM and AFM images of the films. For condition 1, where the Nb and Sn layer thicknesses were 10 nm, several clusters separated by large voids were observed. The grains were not visible on the surface of this film. However, for the other three conditions, the films had surfaces with uniformly distributed grains showing some voids. The film grown with condition 4, where the Nb:Sn thickness ratio was 4:1, showed smaller grains than that for other conditions. The surface roughness for condition 1 was higher due to the film having clusters of larger heights and deeper voids. The RMS roughness for this condition is 66.2±24.3 nm. The large standard deviation originated from the irregular distribution of the clusters and voids. Films with the other three conditions had smoother surfaces. The average roughness was measured to be 31.8±3.2, 31.6±0.9, and 26.8±1.2 nm for conditions 2, 3, and 4, respectively.

The XRD patterns of the films with different Nb thicknesses (10, 20, 30, 40 nm) and constant Sn thickness of 10 nm after annealing at 950 °C for 3 h are compared in Figure 2 (a). All annealed films were polycrystalline $Nb_3Sn$ films showing (200), (210), (222), (320), (321), (400), (420), (421) diffraction orders. The crystallite size and lattice parameter corresponding to $Nb_3Sn$ (210) diffraction peak of the films as a function of Nb:Sn thickness ratio are shown in Figure 1 (b). The crystallite size decreased with increasing Nb layer thickness, which is in agreement with the grain size observed in SEM and AFM. Lattice parameters calculated from the $d$ value of XRD peak are less than lattice parameter of bulk $Nb_3Sn$ (5.290 Å). This is because the Sn composition in the film is less than ideal condition of 25%.

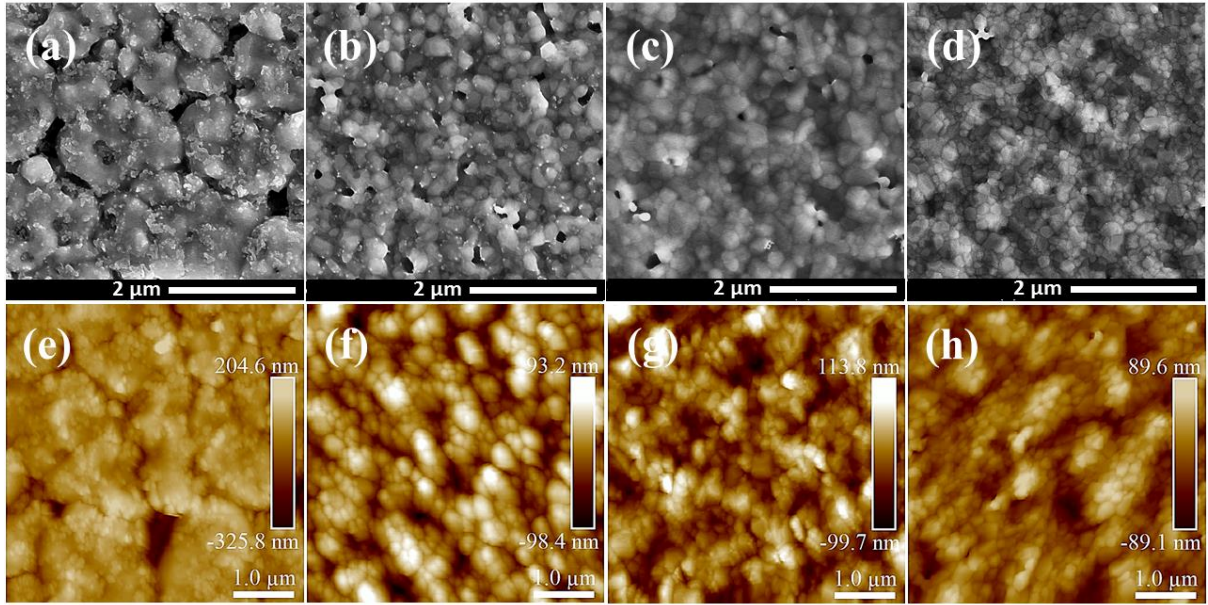

**Figure 1.** SEM and AFM images of the films: (a) and (e) condition 1, (b) and (f) condition 2, (c) and (g) condition 3, (d) and (h) condition 4.

### 3.1.2. *Superconducting properties.*

The resistivity versus temperature is shown in Figure 2 (c). The calculated $T_c$, and $\Delta T_c$ from the graph are shown in Table 2 and Figure 2 (d). All films exhibited good superconducting properties. The highest critical temperature is observed on the film fabricated with condition 1, where the Nb and Sn thicknesses were the same. Better $T_c$ at this condition was obtained due to higher Sn composition on the annealed films. It has been reported that $T_c$ of $Nb_3Sn$ is dependent on the Sn composition of the films [1]. The transition width also became wider with increasing Nb:Sn.

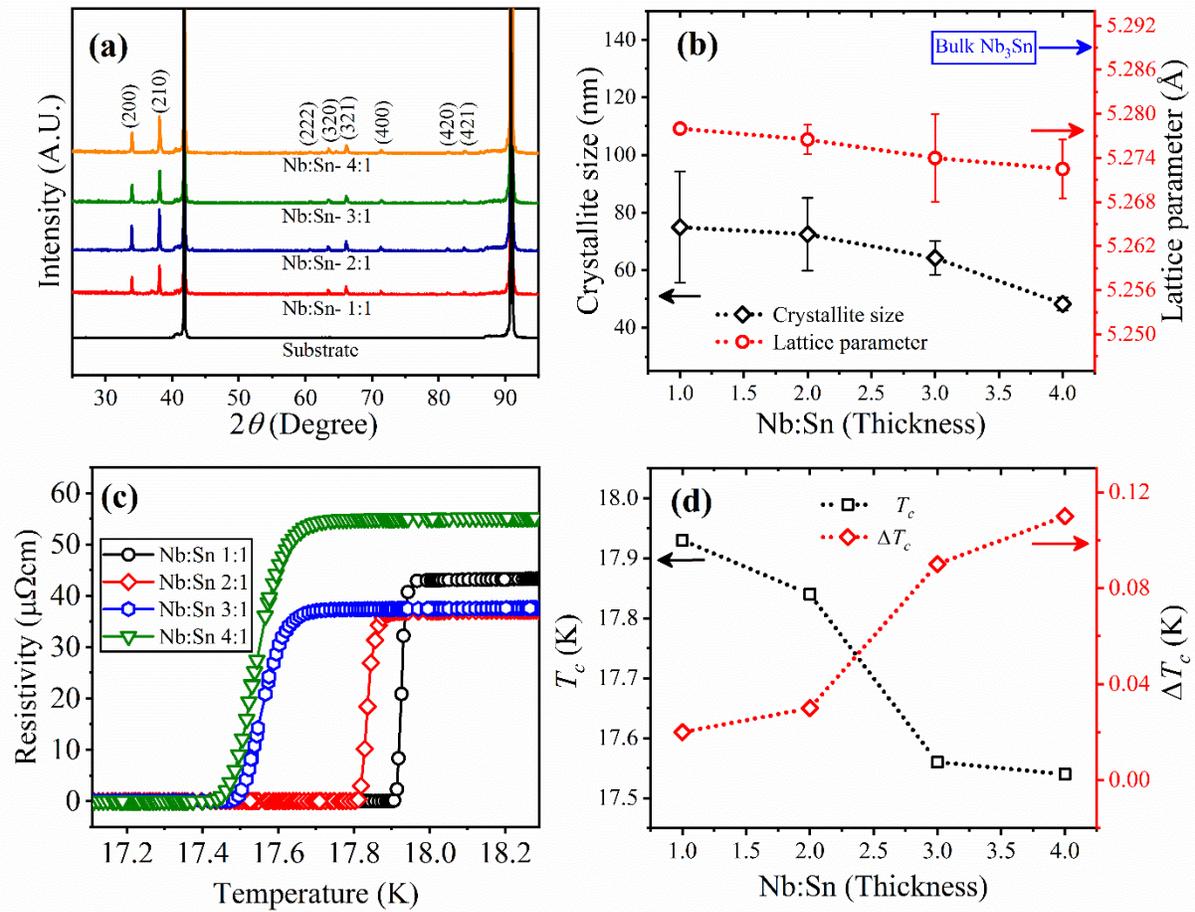

**Figure 2.** (a) X-ray diffraction patterns of the films with conditions 1–4, (b) crystallite size and lattice parameters of the films as a function of Nb:Sn film thickness, (c) resistivity versus temperature of the films, (d) corresponding $T_c$ and $\Delta T_c$ for different Nb:Sn layer thicknesses.

### 3.2. Films with similar stoichiometry

#### 3.2.1. Structural properties.

All annealed films (conditions 4–8) had Sn composition in the range of 21–23%. Figure 3 shows the SEM and AFM images of annealed films of different thicknesses. These films showed uniformly distributed grains with some voids, similar to those fabricated under conditions 1–4. However, for films fabricated under condition 8 (Figure 3 (d) and (h)), the presence of deep voids was not observed. Surface of the film for this condition shows small grains aggregating into clusters. The color pattern on the AFM image of the surface in Figure 3 (h) confirmed that the clusters have different heights, which caused a relatively rough surface. RMS roughness of the films are shown in Table 2. The surface became increasingly rough with increased layer thickness. The lowest roughness of 28.2±2.9 nm was observed for films of condition 5 (Sn layer thickness 5 nm), whereas films of condition 8 (Sn layer thickness 100 nm) had the highest surface roughness of 64.7±3.2 nm.

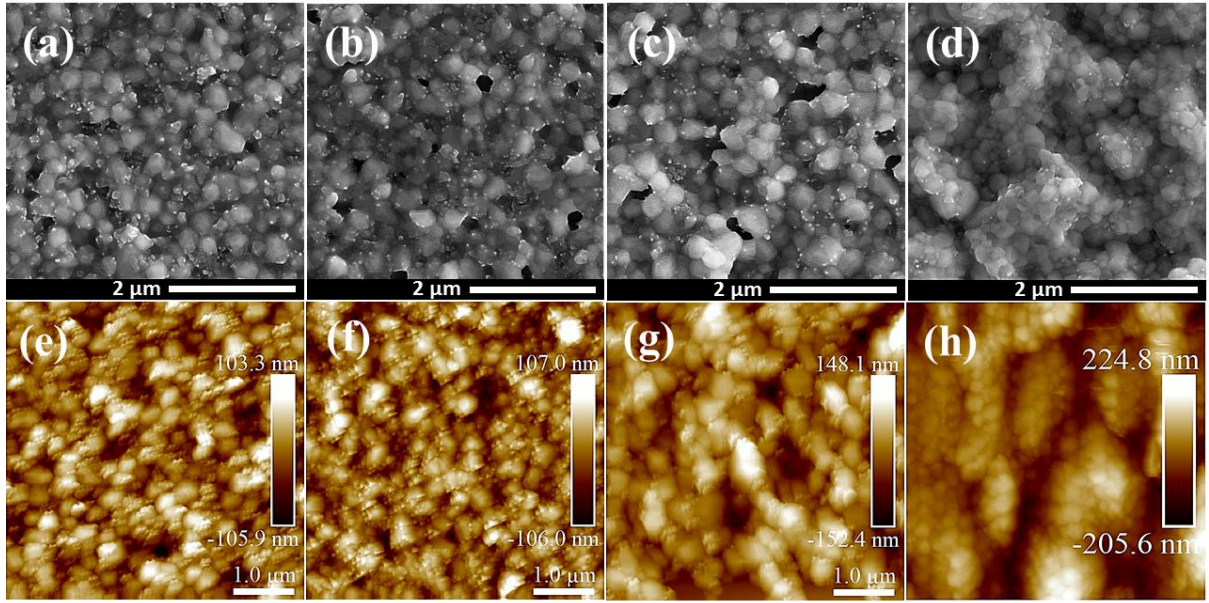

**Figure 3.** SEM and AFM images of the films: (a) and (e) condition 5, (b) and (f) condition 6, (c) and (g) condition 7, (d) and (h) condition 8.

Figure 4 (a) shows the XRD patterns of annealed films with condition 5–8. Similar to the films fabricated under condition 1–4, these films also exhibit diffraction peaks corresponding to $Nb_3Sn$ (200), (210), (222), (320), (321), (400), (420), (421) orders. The lattice parameter and crystallite size of (210) reflection is shown in Figure 4 (b). The lattice parameter and crystallite size correspond to the film fabricated under condition 2 is also plotted. The lattice parameter of the films showed little variation due to the almost similar Sn composition of the films. The crystallite size of the films also did not vary significantly with increased layer thickness. Films for condition 5 had crystallites with an average size of 72.1±5.2 nm, whereas the films for condition 8 had crystallites with an average size of 68.5±2.5 nm.

### 3.2.2. *Superconducting properties.*

The surface resistivity of the films for different Sn layer thicknesses is shown in Figure 4 (c) and the corresponding $T_c$ and $\Delta T_c$ data are shown in Table 2 and Figure 4 (d), respectively. The $T_c$ and $\Delta T_c$ of all films are close. Slight increased $T_c$ is observed on the film for condition 8, however, the $\Delta T_c$ also increased.

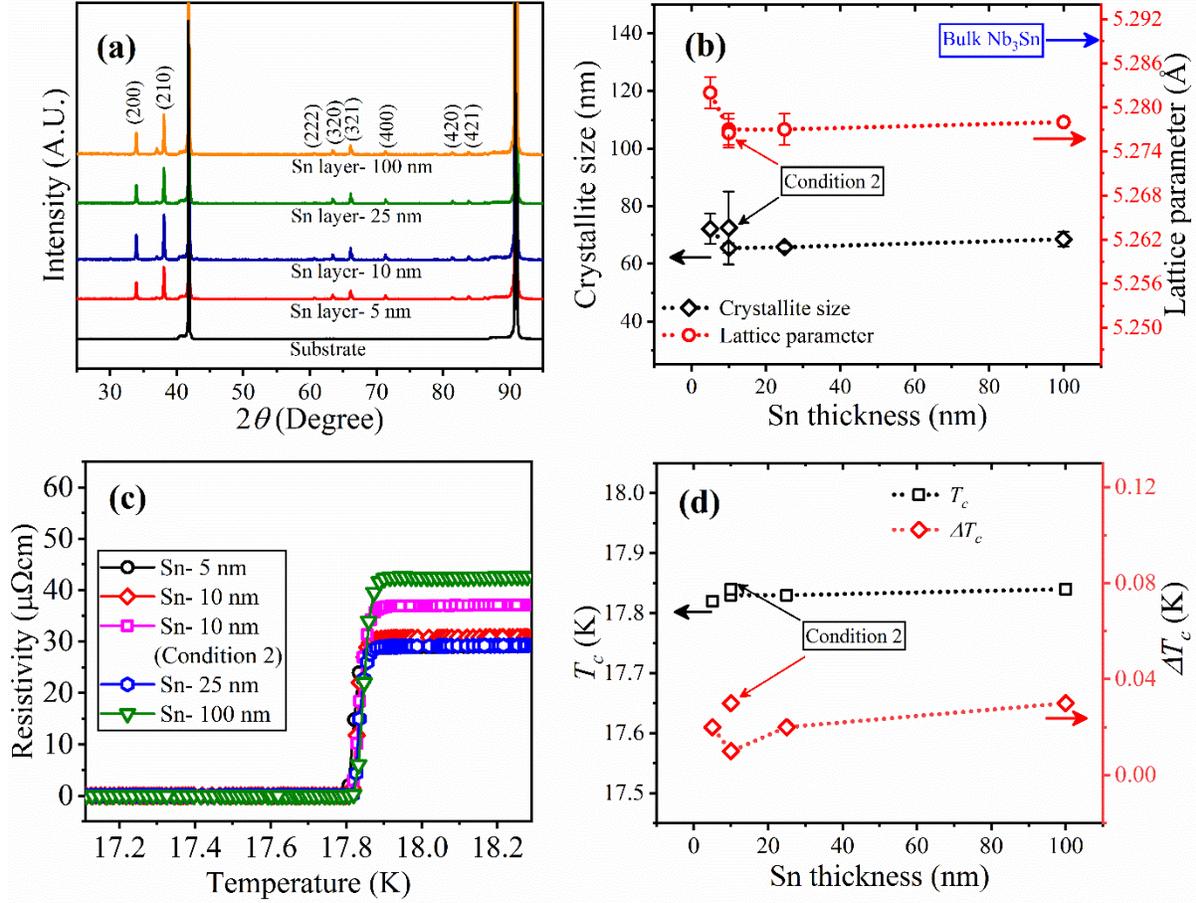

**Figure 4.** (a) X-ray diffraction patterns of the films with condition 5–8, (b) crystallite size and lattice parameters of the films as a function of Sn film thickness, (c) resistivity vs temperature curve of the films, (d) corresponding $T_c$ and $\Delta T_c$ as a function of Sn layer thickness with condition 5–8.

## 4. Discussion

Sn composition plays an important role on the superconducting properties of $Nb_3Sn$. In SRF cavities, $Nb_3Sn$ films coated by conventional Sn vapor diffusion method resulted to Sn deficient patchy regions which degraded the performance of SRF cavities [10, 11]. Therefore, it is important to maintain the stoichiometry of the films after annealing. We have experienced Sn loss for almost all deposition conditions after annealing. This Sn loss resulted from the Sn diffusion and evaporation from the surface. The Sn composition for annealed films obtained from EDS in conditions 3, 4 and 8 are higher than that of as-deposited films. This is due to the limitation of the spatial resolution of EDS. For 15 keV electron, the Anderson-Hasler equation [12] gives the X-ray transmission fraction of 0.68 µm for Nb-Lα lines and 0.76 µm for Sn-Lα lines, which is shorter than the thicknesses of the deposited films. Because of the short X-ray escape depth, layer ordering and the individual layer thickness affects the apparent EDS composition. Films, where the top layer is thick Nb, such as films with condition 3, 4, and 8, appear Nb-rich in EDS measurements. The annealed film showed more Sn due to the uniform distribution of Sn throughout the film after annealing.

The critical temperature of $Nb_3Sn$ varies from 6 to 18.3 K when the Sn composition varies from 17 to 26% [1]. According to the critical temperature versus atomic Sn content plot reported by Godeke [1], the $T_c$ of films with 22% Sn should be ~10–12 K whereas our fabricated films had $T_c$ of ~17.83 K. One possible explanation of higher $T_c$ on our fabricated films is Sn content variation in $Nb_3Sn$. Depth profile of the films can provide the distribution of Sn content throughout the films. Though we reported ToF-SIMS depth profile of the films coated on Nb substrates before [13], we have not observed it for films on sapphire substrates. Off-stoichiometric $Nb_3Sn$ films with high $T_c$ were observed previously on the films grown on sapphire substrate by magnetron sputtering from a single target. Kampwirth *et al.* reported $Nb_3Sn$ film on sapphire substrate with superconducting transition range of 17–17.5 K where atomic Sn composition was ~22% [14]. Recently, Ilyina *et al.* reported $Nb_3Sn$ films with a $T_c$ of 17.5 K on $Al_2O_3$ substrate whereas films of same condition on copper substrates resulted to reduced $T_c$ of 14.7 K [7].

The surface morphology of the films is affected by the coating parameters. It is possible to change the surface roughness varying the thickness of Nb and Sn layers. The grain sizes of the films were also varied when Nb:Sn ratio was varied. The cross-section image of these films could give more detailed idea about the grain structure near the surface as well as the interface. The technique has been used on vapor-diffused samples [15], but not yet on the present films.

## 5. Conclusion

$Nb_3Sn$ films were fabricated by sputtering multilayers of Nb and Sn with varied layer thicknesses followed by annealing. Throughout annealing, evaporation of Sn from the surface and growth of $Nb_3Sn$ by diffusion of Sn into Nb surface occur simultaneously. Sn loss is the outcome of the competition of these two processes. Sn loss depends on annealing parameters (annealing temperature and annealing time). For annealing at 950 °C for 3 h, Sn composition were 20–23% for all coating conditions. The surface morphology and the roughness of the films varied with the thickness of the layers. For different Nb:Sn thicknesses, the grain size and surface roughness decreased with increasing thickness ratio. For different layer thicknesses with constant thickness ratio, the surface roughness increased with increasing thickness of the layers, whereas the grain size did not vary significantly. All films showed good superconducting properties with a superconducting critical temperature $T_c$ up to 17.93 K.

**Acknowledgments**

This material is based upon work supported by the U.S. Department of Energy, Office of Science, Office of Nuclear Physics under Contract No. DE-AC05-06OR23177. The authors acknowledge Michael J. Kelley, Gianluigi Ciovati for their suggestions and Joshua Spradlin for his help with $T_c$ measurement.